\documentclass[10pt,conference]{IEEEtran}
\usepackage{dblfloatfix}
\usepackage{booktabs}
\usepackage{array}
\usepackage{adjustbox}
\usepackage{balance}
\usepackage{amsmath,amssymb}
\usepackage{xcolor}
\usepackage{tikz}
\usetikzlibrary{positioning,fit,backgrounds,calc,shapes.geometric}
\usepackage[most]{tcolorbox}
\usepackage{graphicx}
\usepackage{multirow}
\usepackage{wrapfig}
\usepackage[hidelinks]{hyperref}
\newcolumntype{C}[1]{>{\centering\arraybackslash}p{#1}}
\newcommand{\dtoh}{\ensuremath{d2h}}
\newcommand{\ezr}{EZR}
\newcommand{\de}{DE}
\definecolor{boxgray}{HTML}{F4F4F6}
\definecolor{boxrule}{HTML}{8C1515}
\definecolor{famEvo}{HTML}{FC8D62}
\definecolor{famSur}{HTML}{66C2A5}
\definecolor{famGeo}{HTML}{E78AC3}
\definecolor{famTra}{HTML}{8DA0CB}
\definecolor{famRnd}{HTML}{BDBDBD}
\newtcolorbox{finding}[1][]{
  enhanced, breakable, colback=boxgray, colframe=boxrule,
  boxrule=0.4pt, left=6pt, right=6pt, top=5pt, bottom=5pt,
  leftrule=3pt, arc=1pt, #1}
\begin{document}
\title{Which Optimizer, At What Budget? \\A Tournament of   Optimizers
for Search-Based SE}
\author{
\IEEEauthorblockN{Kishan Kumar Ganguly and Tim Menzies}
\IEEEauthorblockA{
North Carolina State University\\
kgangul@ncsu.edu, timm@ieee.org
}
}
\maketitle

\begin{abstract}
Configuring and tuning modern software is unavoidable, expensive, and error-prone: a single system can expose hundreds of interacting options, and scoring one setting can mean a full build or test run. The standard response is automated optimization, but the number of available optimizers is large and growing. And some of the guidance for selecting among them is misleading: NSGA-II, for example, is widely recommended, yet other algorithms reach the same results using only 1/20th as many evaluations. 

To help practitioners make better choices about tools to configure their systems, we cluster 20 optimizers, based on  six assumptions about the data. Next, we run a tournament across those optimizers, using 106 SE optimization tasks at four labeling budgets (taking 14,000+ CPU hours). We find that no optimizer wins outright. The best one migrates with the budget (from a geometric active learner when labels are scarce to differential evolution when labels are plentiful) so a winner ``crowned'' at one budget is wrong at another on up to half our tasks. 

Running such a tournament for every new domain is impractical due to its CPU cost. Fortunately, we find that those 14,000 hours can be replaced by a table lookup over two cheap-to-obtain task attributes (plus the labeling budget). Predictions from this table tie or beat a hindsight oracle on $\approx 75\%$ of held-out tasks. 

To support open science, our tournament and replication package are open-sourced for SBSE researchers and practitioners at \url{https://github.com/KKGanguly/OptimizerTournament}.
\end{abstract}

\maketitle
\begin{table*}[t]
\centering
\begin{minipage}[t]{0.44\textwidth}
\centering
\caption{  \textbf{106} MOOT SE tasks used here,
grouped by domain. ``x/y'' is inputs/outputs (decisions/goals). Tasks with
$y\!\ge\!2$ goals are  multi-objective. The set is the SE core of MOOT;
non-SE tasks (sales, finance, generic ML) are excluded.}
\label{tab:datasets}
\footnotesize
\setlength{\tabcolsep}{2pt}
\renewcommand{\arraystretch}{1.05}
\begin{adjustbox}{max width=\linewidth}
\begin{tabular}{@{}rp{2.45cm}>{\raggedright\arraybackslash}p{2.7cm}cc@{}}
\toprule
\textbf{\#} & \textbf{Domain / type} & \textbf{Example tasks} & \textbf{x/y} & \textbf{\# rows}\\
\midrule
25 & Specific software config. & \texttt{SS-A}\,$\ldots$\,\texttt{SS-X}, \texttt{billing10k} & 3--88 / 2--3 & 0.2k--86k\\
12 & Performance (systems) & \texttt{7z, BDBC, LLVM, Postgre\-SQL, x264, redis}\,$\ldots$ & 9--35 / 1 & 0.9k--167k\\
11 & Cloud / system tuning & \texttt{Apache, SQL, HSMGP}, \texttt{rs/sol/wc}\,$\ldots$ & 3--39 / 1 & 0.2k--4.7k\\
35 & Project health & \texttt{Health-\{Issues, PRs, Commits\}} & 5 / 2--3 & 10k\\
11 & Feature models & \texttt{Scrum\{1,10,100\}k}, \texttt{FFM-*}, \texttt{FM-*} & 124--1044 / 3 & 1k--100k\\
10 & Process / cost models & \texttt{nasa93dem, coc1000, POM3a--d, XOMO} & 9--27 / 3--5 & 93--20k\\
~2 & Testing & \texttt{test120, test600} & 9 / 1 & 5.2k\\
\midrule
\textbf{106} & \textbf{Total} & \multicolumn{3}{l}{\emph{spanning 7 SE domains; 25 single- and 81 multi-objective}}\\
\bottomrule
\end{tabular}
\end{adjustbox}
\\[2pt]
{\scriptsize Sources:
\cite{Amiraliminimaldata,menzies2025the,chen2026promisetune,senthilkumar2024can,lusstosa2025less,lustosa2024learning,nair2018finding}.}
\end{minipage}\hfill
\begin{minipage}[t]{0.54\textwidth}
\centering
\caption{Inventory of the optimizers placed in the tournament. For each we give the assumption it bets on, its role, the originating reference, representative
\emph{software-engineering} uses, and the venue of those uses. This inventory is an
artifact of the literature review (Section~\ref{sec:related}) and the menu the
tournament of Section~\ref{sec:method}.}
\label{tab:inventory}
\footnotesize
\setlength{\tabcolsep}{5pt}
\renewcommand{\arraystretch}{1.12}
\begin{adjustbox}{max width=\linewidth}
\begin{tabular}{@{}llll l l@{}}
\toprule
\textbf{Optimizer} & \textbf{Bet} & \textbf{Role} & \textbf{Orig.} & \textbf{SE usage} & \textbf{Venue of SE uses}\\
\midrule
Hill Climbing      & A1 & greedy trajectory    & \cite{russell2009artificial}            & \cite{harman2009theoretical,harman2009search,he2021pyart}                  & TSE, ACM CSUR, ICSE\\
Simulated Annealing& A1 & annealed trajectory  & \cite{kirkpatrick1983optimization}      & \cite{harman2012search,wang2011search}                   & FOSE, ASE\\
(1+1)-ES           & A1 & self-adaptive step trajectory        & \cite{rechenberg1973evolutionsstrategie} & \cite{arcuri2018test}                              & IST\\
Iterated Local Search& A2& restart escape      & \cite{lourenco2003iterated}             & \cite{sabar2018genetic}                                    & GECCO\\
Tabu Search        & A2 & memory escape        & \cite{glover1989tabu}                   & \cite{diaz2008tabu,lin2015tca,agrawal2019dodge}           & Comput.\ \& OR, IST, TSE\\
Genetic Algorithm  & A3 & crossover population & \cite{holland1975adaptation}            & \cite{legoues2011genprog,zhang2018empirical}  & TSE, TOSEM, ICSE-W\\
EDA                & A3 & distribution model   & \cite{muhlenbein1996recombination}      & \cite{wei2024test}                                    & GECCO\\
PSO                & A3 & velocity swarm       & \cite{kennedy1995particle}              & \cite{lee2023learning}                                    & ICSE\\
DE                 & A3 & difference-vector search & \cite{storn1997differential}       & \cite{fu2017easy}                        & ESEC/FSE\\
SMAC               & A5 & RF surrogate BO      & \cite{hutter2011sequential}             & \cite{ganguly2026lowgodatalightse,chen2026promisetune}  & FSE, ICSE\\
TPE                & A5 & density estimation   & \cite{bergstra2011algorithms}           & \cite{chen2021efficient,ganguly2026lowgodatalightse}                 & ICSE, FSE\\
LINE (kpp)         & A7 & centroid sampling    & \cite{ganguly2026lowgodatalightse}               & \cite{ganguly2026lowgodatalightse, Amiraliminimaldata}                                 & FSE, JSS\\
SWAY               & A7 & distance bisection   & \cite{chen2019sampling}                 & \cite{chen2019sampling,chen2025accuracy}                  & TSE, TSE\\
EZR                & A7 & proximity active learning& \cite{Amiraliminimaldata}              & \cite{Amiraliminimaldata,ganguly2026lowgodatalightse}             & JSS, FSE\\
DODGE              & A7 & $\epsilon$-pruning   & \cite{agrawal2019dodge}                 & \cite{agrawal2019dodge}                                    & TSE\\
Random Search      & floor & blind probing     & \cite{bergstra2012random}               & \cite{chen2019sampling,Amiraliminimaldata, ganguly2026lowgodatalightse, chen2026promisetune}                & TSE, JSS, FSE, ICSE\\
NSGA-II            & A4 & dominance sort       & \cite{deb2002fast}                      & \cite{mkaouer2015many,matnei2016multi}                    & TOSEM, JSERD\\
SPEA2              & A4 & strength archive     & \cite{zitzler2001spea2}                 & \cite{matnei2016multi}                                     & JSERD\\
SMS-EMOA           & A4 & hypervolume select   & \cite{beume2007sms}                     & \cite{ni2019empirical}                                 & JSS\\
MOEA/D             & A4 & decomposition        & \cite{zhang2007moead}                   & \cite{mkaouer2015many}                       & TOSEM\\
\bottomrule
\end{tabular}
\end{adjustbox}
\end{minipage}
\end{table*}
\begin{figure*}[!b]
  \centering
  \includegraphics[width=0.6\textwidth]{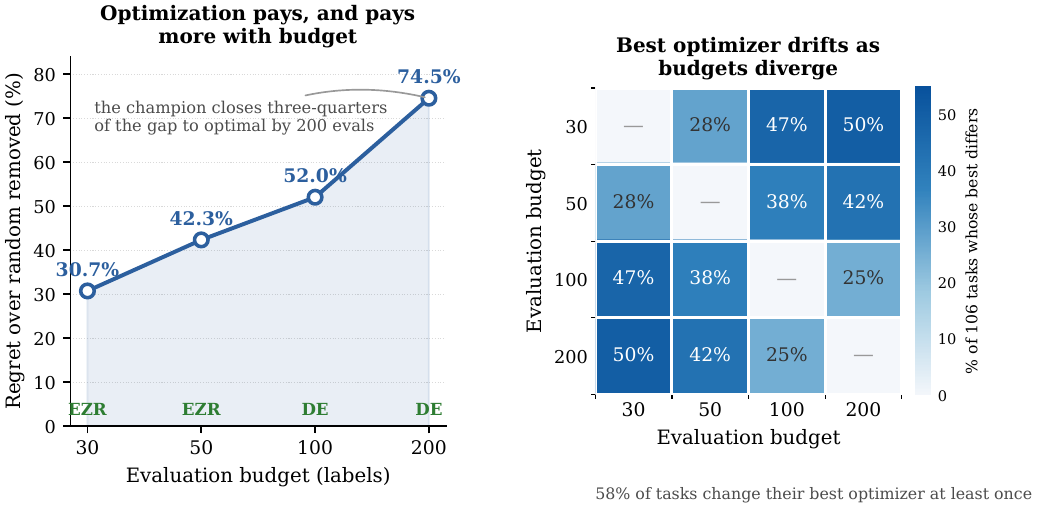}
\caption{Optimization can be very effective for SE tasks.
Left panel: In 106 MOOT SE tasks,
reasoning on just 30 labels already closes \textbf{31\%} of the
gap between a random guess and the best setting;  and 200 labels does even better (reaches to \textbf{75\%} of
the optimum). 
\textbf{Right panel:} the best optimizer
changes with budget (58\% of tasks switch at least once so no single choice is
stable across all budgets).
For this reason, this paper
builds a selection rule that can guide practitioners on when to use
which optimizer.}
  \label{fig:stakes}
\end{figure*}


\section{Introduction}
\label{sec:intro}

Configuring and tuning modern software is unavoidable and expensive. A modern system
stacks languages, libraries, compilers, and deployment pipelines, each exposing its own
flags and defaults: one open-source database used in this study carries 460 binary
options, creating a configuration space of $2^{460}$, larger than the number of stars in the
sky~\cite{Amiraliminimaldata}. Exhaustive search is impractical at this
scale, so the field turns to \emph{search-based software engineering}
(SBSE) to formulate this as an optimization problem. Under this paradigm,a fitness function scores candidate configurations and a search algorithm
walks the space looking for good ones~\cite{harman2009search,harman2012search}.

We study black-box optimization: reasoning about a system from inputs and outputs alone.
It is the workhorse of SE configuration, used to tune compilers and databases~\cite{chen2023compiler,van2017automatic}, configure product lines and cloud
systems~\cite{sayyad2013value,zhang2019end}, and generate adversarial
tests~\cite{huang2025iterative}. Over 100 such tasks are documented in the
MOOT repository (Table~\ref{tab:datasets}), representing more than a decade of SBSE
work~\cite{menzies2025the}.

A practitioner who reaches for these tools confronts an overwhelming set of choices.
Table~\ref{tab:inventory} lists 20 optimizers drawn from the recent SE literature, and
the No-Free-Lunch theorem guarantees that none of them wins
everywhere~\cite{wolpert1997no,ma2025toward}. Which optimizer suits which task is, today,
largely guesswork.

The \emph{labeling budget} compounds the problem. A single SE fitness evaluation can be
costly: it may run an entire test suite for program repair~\cite{legoues2011genprog},
execute thousands of unit tests to score LLM-generated code~\cite{qiu2025efficient}, or
profile system-level time and energy. Agentic and LLM-driven workflows push this cost
higher. Still, An engineer may afford only dozens of evaluations, and the best optimizer
at thirty evaluations need not be the best at two hundred. This exposes a gap:

\begin{finding}
\textbf{The gap.} SBSE has no systematic way to organize optimizers by the assumptions
they encode, and no cheap, budget-aware rule for choosing one on a given task.
\vspace{-0.2cm}
\end{finding}

We close this gap with an assumption-indexed tournament. Reading the SBSE literature, we
map each dominant search assumption to one representative optimizer, then run 20
black-box optimizers as bets on six of our seven assumptions (Table~\ref{tab:inventory}) over 106
MOOT tasks (Table~\ref{tab:datasets}) at four labeling budgets, repeated 20 times:
roughly 14{,}000 CPU hours
(the largest such SBSE study we are aware of).

We organize the results around four questions:
\begin{itemize}
\item \textbf{RQ1: Is optimization worth doing?} Yes, even a few dozen labels close
\textbf{31--75\%} of the gap between an uninformed choice and the best setting.
\item \textbf{RQ2: Does the budget change which optimizer wins?} Yes, the winner
migrates from a geometric active learner (EZR~\cite{ganguly2026lowgodatalightse}) when
labels are scarce to differential evolution (DE~\cite{storn1997differential}) when they
are plentiful. And an optimizer selected
at one budget falls outside the target budget's top Scott-Knott tier on up to
\textbf{50\%} of tasks.
\item \textbf{RQ3:  Do these optimizers really differ?}
Yes: single-objective search beats multi-objective methods at equal budget. NSGA-II needs 1000 samples to reach what EZR reaches in 200.
\item \textbf{RQ4: Can we cheaply predict the winner?} Yes, a guide based on  two cheap-to-obtain task
attributes plus the budget ties or beats a hindsight oracle on \textbf{$\approx75\%$} of
held-out tasks, far above the \textbf{$\approx45\%$} ceiling of expensive instance clustering features.
\end{itemize}

This paper contributes:
\begin{itemize}
\item A catalog of black-box optimizers organized by seven assumptions, read from the literature and used as an organizing device 
(Section~\ref{sec:related});
\item An assumption-indexed, budget-indexed tournament over 20 optimizers and 106 SE
tasks, a reusable rig for SBSE experimentation (Section~\ref{sec:method});
\item An easy-to-use, validated, budget-aware selection guide keyed on two cheap-to-obtain 
attributes (plus the labeling budget) (Section~\ref{sec:rq4}).
\item An open source reproduction package  at \url{https://github.com/KKGanguly/OptimizerTournament}.
\end{itemize}

\section{Motivation}
\label{sec:motivation}

Optimization is very effective for SE tasks, even when only a few examples can be
labeled. The space of possible configurations
is  enormous: one system we studied  has 460 binary flags, a
space of $2^{460}$ options
(which means there are more configurations than there are stars in the
observable universe~\cite{Amiraliminimaldata}). An engineer cannot try them all,
and does not have to. Take the honest baseline
(pick one configuration at
random), then spend just \textbf{30 labels} (30 builds, or 30 benchmark runs) and
let an optimizer choose what to try. 
Across our 106 tasks that small amount of data closes 31\% of the gap between the random guess and the best setting. With 200 labels, still a vanishing fraction of spaces as large as $2^{460}$, the methods of this paper close 75\% of that gap (Figure~\ref{fig:stakes}, left panel).

Labeling very few examples works for even very
large problems. Returning to that example with $2^{460}$ options, with just \textbf{50 labels}
(0.5\% of the data) the methods used in this
paper can reach \textbf{80\% of
optimal}~\cite{Amiraliminimaldata}. 
Hence we say: 
\begin{quote}
{\em The  open question in SE optimization
is never \emph{whether} to optimize, only \emph{which} optimizer earns those few
dozen labels.}
\end{quote}
But there is a catch. 
The right panel of Figure~\ref{fig:stakes}
warns that deciding which optimizer to select is  a nuanced task. That figure compares, for each pair of budgets $x$ and $y$, the percentage of tasks whose best optimizer at $x$ differs from their best at $y$. Note
that, very often, the ``best'' optimizer   changed
when we can access more labels. For this reason, this
paper builds a selection rule that can guide practitioners on when to use which optimizer.

The rest of this section offers
other motivation for this
exploration of
optimization for SE tasks.

\subsection{Why Evaluate Optimizers on SE Data?}
SE optimization
has certain features which
require special kinds of
optimizers~\cite{agrawal2022simpler}. 
Every ``if'' statement divides
the internal state space of a program into different regions, so there is
no single gradient to follow.
This means that one cannot (e.g.)  differentiate a \texttt{for} loop.
The search space is marked by
discrete options (e.g. boolean configuration options) and is
interaction-heavy rather
than smooth~\cite{aleti2017analysing,albunian2020causes}.  Limits are hard
constraints, not soft penalties (one byte over a memory cap is a failure, not a
gradient).
And evaluating any candidate means building and running it \cite{harman2012search}. 
Together, these make SE a budgeted, complex, black-box problem, the kind of problem where the 
No-Free-Lunch bites hardest~\cite{wolpert1997no}: the optimizer that wins in ML may
not win here \cite{agrawal2022simpler}. 

\begin{finding}
\textbf{SE needs its own optimizers.} SE optimization is budgeted, rugged, and
black-box, so methods proven on smooth ML losses may not work in our domain. Choosing the
optimizer is itself the problem, and No-Free-Lunch guarantees no single choice is
always right.
\end{finding}


\definecolor{soFill}{RGB}{233,233,250}
\definecolor{soDraw}{RGB}{120,120,200}
\definecolor{moFill}{RGB}{253,232,205}
\definecolor{moDraw}{RGB}{220,150,70}
\definecolor{stageFill}{RGB}{210,212,240}
\definecolor{grandFill}{RGB}{248,210,214}
\definecolor{grandDraw}{RGB}{200,90,100}
\definecolor{winGreen}{RGB}{30,120,40}
\definecolor{t1blue}{RGB}{70,80,180}
\definecolor{t2orange}{RGB}{200,110,30}

\providecommand{\wtick}{}
\renewcommand{\wtick}{%
  \raisebox{-0.15ex}{%
    \begin{tikzpicture}[scale=0.075]
      \draw[winGreen,line width=1.5pt,line cap=round,line join=round]
        (0,1)--(0.85,0)--(2.5,2.4);
    \end{tikzpicture}%
  }%
}

\providecommand{\BR}{}
\renewcommand{\BR}[4]{%
  \draw[conn]
  let \p1=(#1.east), \p2=(#2.east), \p3=(#3.west) in
    (\p1) -- (#4,\y1)
    (\p2) -- (#4,\y2)
    (#4,\y1) -- (#4,\y2)
    (#4,\y3) -- (\p3);
}

\providecommand{\BRthree}{}
\renewcommand{\BRthree}[5]{%
  \draw[conn]
  let \p1=(#1.east), \p2=(#2.east), \p3=(#3.east), \p4=(#4.west) in
    (\p1) -- (#5,\y1)
    (\p2) -- (#5,\y2)
    (\p3) -- (#5,\y3)
    (#5,\y1) -- (#5,\y3)
    (#5,\y4) -- (\p4);
}

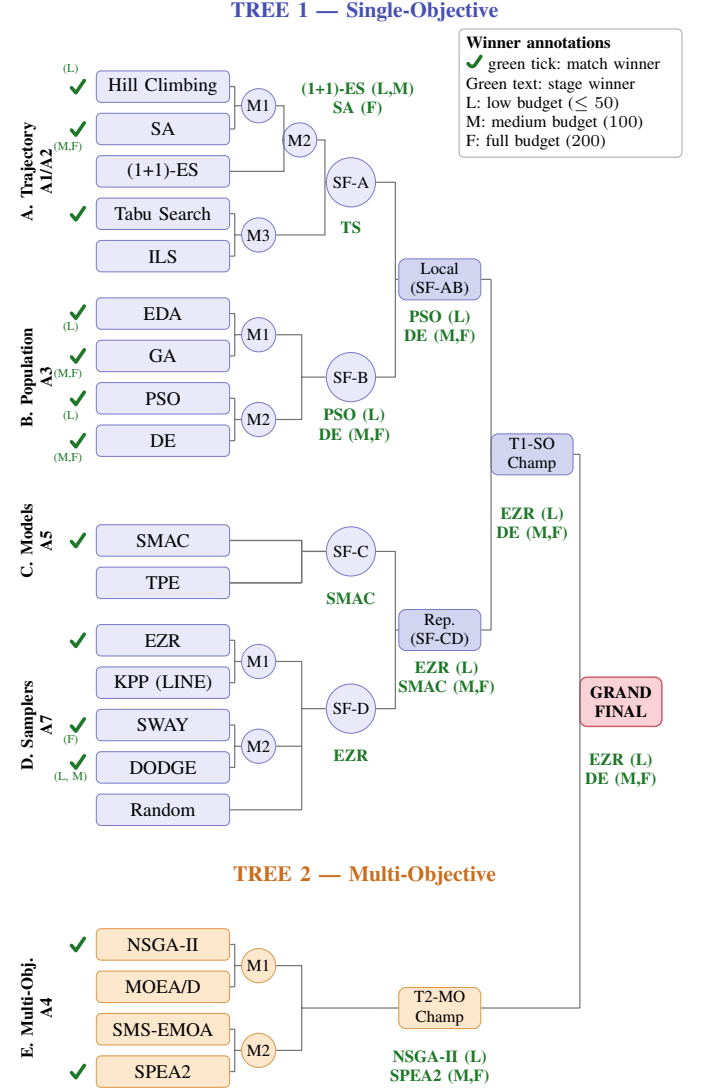
\begin{figure}[!b]
\centering
\resizebox{\columnwidth}{!}{%
\begin{tikzpicture}[
  x=1cm,y=1cm,font=\footnotesize,
  leaf/.style={draw=soDraw,fill=soFill,rounded corners=2pt,
               minimum width=1.95cm,minimum height=0.46cm,
               inner sep=1pt,align=center},
  moleaf/.style={leaf,draw=moDraw,fill=moFill},
  mn/.style={draw=soDraw,fill=soFill,circle,
             minimum size=0.50cm,inner sep=0pt,font=\scriptsize},
  momn/.style={mn,draw=moDraw,fill=moFill},
  sfn/.style={draw=soDraw,fill=soFill,circle,
              minimum size=0.72cm,inner sep=0pt,
              font=\scriptsize,align=center},
  stage/.style={draw=soDraw,fill=stageFill,rounded corners=3pt,
                minimum width=1.20cm,minimum height=0.60cm,
                inner sep=1pt,font=\scriptsize,align=center},
  mostage/.style={stage,draw=moDraw,fill=moFill},
  grand/.style={draw=grandDraw,fill=grandFill,rounded corners=3pt,
                line width=0.6pt,minimum width=1.20cm,
                minimum height=0.72cm,inner sep=1pt,
                font=\scriptsize\bfseries,align=center},
  win/.style={text=winGreen,font=\scriptsize\bfseries,
              align=center,inner sep=1pt},
  fam/.style={rotate=90,font=\scriptsize\bfseries,
              align=center,anchor=center},
  conn/.style={draw=black!55,line width=0.5pt},
]

\def\xl{1.05}
\def\xm{2.45}
\def\xsf{3.80}
\def\xab{5.10}
\def\xt{6.45}
\def\xg{7.75}

\def\jone{2.10}
\def\jtwo{3.08}
\def\jthree{4.45}
\def\jfour{5.85}
\def\jfive{7.15}

\node[leaf] (HC)  at (\xl, 0.00){Hill Climbing};
\node[leaf] (SA)  at (\xl,-0.62){SA};
\node[leaf] (ES)  at (\xl,-1.24){(1+1)-ES};
\node[leaf] (TS)  at (\xl,-1.86){Tabu Search};
\node[leaf] (ILS) at (\xl,-2.48){ILS};

\node[leaf] (EDA) at (\xl,-3.32){EDA};
\node[leaf] (GA)  at (\xl,-3.94){GA};
\node[leaf] (PSO) at (\xl,-4.56){PSO};
\node[leaf] (DE)  at (\xl,-5.18){DE};

\node[leaf] (SMAC) at (\xl,-6.64){SMAC};
\node[leaf] (TPE)  at (\xl,-7.26){TPE};

\node[leaf] (EZR)   at (\xl,-8.10){EZR};
\node[leaf] (KPP) at (\xl,-8.72){KPP (LINE)};
\node[leaf] (SWAY)  at (\xl,-9.34){SWAY};
\node[leaf] (DODGE)   at (\xl,-9.96){DODGE};
\node[leaf] (RND)   at (\xl,-10.58){Random};

\node[anchor=east] at (0.05, 0.00){\wtick};
\node[win,font=\tiny] at (-0.30, 0.26){(L)};

\node[anchor=east] at (0.05,-0.62){\wtick};
\node[win,font=\tiny] at (-0.34,-0.88){(M,F)};

\node[anchor=east] at (0.05,-1.86){\wtick};

\node[anchor=east] at (0.05,-3.30){\wtick};
\node[win,font=\tiny] at (-0.30,-3.52){(L)};
\node[anchor=east] at (0.05,-3.94){\wtick};
\node[win,font=\tiny] at (-0.34,-4.20){(M,F)};

\node[anchor=east] at (0.05,-4.56){\wtick};
\node[win,font=\tiny] at (-0.30,-4.82){(L)};

\node[anchor=east] at (0.05,-5.18){\wtick};
\node[win,font=\tiny] at (-0.34,-5.44){(M,F)};

\node[anchor=east] at (0.05,-6.64){\wtick};

\node[anchor=east] at (0.05,-8.10){\wtick};
\node[anchor=east] at (0.05,-9.34){\wtick};
\node[win,font=\tiny] at (-0.30,-9.56){(F)};
\node[anchor=east] at (0.05,-9.88){\wtick};
\node[win,font=\tiny] at (-0.30,-10.10){(L, M)};

\node[mn] (MA1) at (\xm,-0.28){M1};
\node[mn] (MA2) at (\xm+0.60,-0.78){M2};
\node[mn] (MA3) at (\xm,-2.17){M3};

\node[mn] (MB1) at (\xm,-3.63){M1};
\node[mn] (MB2) at (\xm,-4.87){M2};

\node[mn] (MD1) at (\xm,-8.41){M1};
\node[mn] (MD2) at (\xm,-9.65){M2};

\node[sfn] (SFA) at (\xsf,-1.40){SF-A};
\node[sfn] (SFB) at (\xsf,-4.25){SF-B};
\node[sfn] (SFC) at (\xsf,-6.80){SF-C};
\node[sfn] (SFD) at (\xsf,-9.10){SF-D};

\node[stage] (SFAB) at (\xab,-2.82){Local\\(SF-AB)};
\node[stage] (SFCD) at (\xab,-7.95){Rep.\\(SF-CD)};
\node[stage] (T1)   at (\xt,-5.38){T1-SO\\Champ};
\node[grand] (GF)   at (\xg,-9.00){GRAND\\FINAL};

\node[win] at (\xm+1.45,-0.18){(1+1)-ES (L,M)\\SA (F)};
\node[win] at (\xsf,-2.06){TS};

\node[win] at (\xsf+0.05,-4.96){PSO (L)\\DE (M,F)};
\node[win] at (\xsf,-7.48){SMAC};
\node[win] at (\xsf,-9.78){EZR};

\node[win] at (\xab,-3.52){PSO (L)\\DE (M,F)};
\node[win] at (\xm+2.75,-8.65){EZR (L)\\SMAC (M,F)};

\node[win] at (\xt,-6.40){EZR (L)\\DE (M,F)};
\node[win] at (\xg,-10.00){EZR (L)\\DE (M,F)};

\BR{HC}{SA}{MA1}{\jone}

\def\jachain{2.82}
\draw[conn] (MA1.east) -- (\jachain,-0.28) -- (\jachain,-0.78) -- (MA2.west);
\draw[conn] (ES.east)  -- (\jachain,-1.24) -- (\jachain,-0.78);

\BR{TS}{ILS}{MA3}{\jone}

\BR{EDA}{GA}{MB1}{\jone}
\BR{PSO}{DE}{MB2}{\jone}

\BR{SMAC}{TPE}{SFC}{\jtwo}

\BR{EZR}{KPP}{MD1}{\jone}
\BR{SWAY}{DODGE}{MD2}{\jone}

\def\jaSF{3.42}
\draw[conn] (MA2.east) -- (\jaSF,-0.78) -- (\jaSF,-1.40) -- (SFA.west);
\draw[conn] (MA3.east) -- (\jaSF,-2.17) -- (\jaSF,-1.40);

\BR{MB1}{MB2}{SFB}{\jtwo}

\BR{SMAC}{TPE}{SFC}{\jtwo}

\BRthree{MD1}{MD2}{RND}{SFD}{\jtwo}

\BR{SFA}{SFB}{SFAB}{\jthree}
\BR{SFC}{SFD}{SFCD}{\jthree}

\BR{SFAB}{SFCD}{T1}{\jfour}

\node[moleaf] (NSGA)  at (\xl,-12.55){NSGA-II};
\node[moleaf] (MOEAD) at (\xl,-13.17){MOEA/D};
\node[moleaf] (SMS)   at (\xl,-13.79){SMS-EMOA};
\node[moleaf] (SPEA)  at (\xl,-14.41){SPEA2};

\node[anchor=east] at (0.05,-12.55){\wtick};
\node[anchor=east] at (0.05,-14.41){\wtick};

\node[momn] (M21) at (\xm,-12.86){M1};
\node[momn] (M22) at (\xm,-14.10){M2};

\node[mostage] (T2) at (\xab,-13.48){T2-MO\\Champ};
\node[win] at (\xab,-14.35){NSGA-II (L)\\SPEA2 (M,F)};

\BR{NSGA}{MOEAD}{M21}{\jone}
\BR{SMS}{SPEA}{M22}{\jone}
\BR{M21}{M22}{T2}{\jtwo}

\BR{T1}{T2}{GF}{\jfive}

\node[fam] at (-0.80,-1.24){A. Trajectory\\\scriptsize A1/A2};
\node[fam] at (-0.80,-4.25){B. Population\\\scriptsize A3};
\node[fam] at (-0.80,-6.64){C. Models\\\scriptsize A5};
\node[fam] at (-0.80,-9.34){D. Samplers\\\scriptsize A7};
\node[fam] at (-0.80,-13.48){E. Multi-Obj.\\\scriptsize A4};

\node[font=\small\bfseries,text=t1blue] at (4.0,1.10)
  {TREE 1 --- Single-Objective};
\node[font=\small\bfseries,text=t2orange] at (4.0,-11.55)
  {TREE 2 --- Multi-Objective};

\node[
  draw=black!35,
  fill=white,
  rounded corners=2pt,
  align=left,
  font=\scriptsize,
  text width=3.05cm,
  inner sep=3pt,
  anchor=north east
] at (8.65,0.85){%
  \textbf{Winner annotations}\\[1pt]
  \wtick~green tick: match winner\\
  Green text: stage winner\\
  L: low budget ($\leq 50$)\\
  M: medium budget ($100$)\\
  F: full budget ($200$)\\
};

\end{tikzpicture}}
\caption{The assumption-indexed tournament tree (20 optimizers). Leaves are
black-box optimizers, colored by family; each internal node is a head-to-head
match isolating one landscape assumption. TREE-1 ranks the single-objective
methods through four assumption branches: Trajectory (A1/A2, with the
continuity chain Hill\,Climbing $\to$ SA $\to$ (1+1)-ES feeding the A-final
against the Tabu/ILS diversification winner), Population (A3, including DE),
Models (A5, comparing SMAC and TPE), and Samplers (A7). Their champions meet in two
semifinals, Local and Representation, to give the T1-SO champion. TREE-2 ranks
the multi-objective methods (A4) into the T2-MO champion, and the Grand Final
compares the two. Green labels give each match's budget-tagged winner
(L$\leq$50, M$=$100, F$=$200). The single-objective champion migrates from EZR
at tight budgets to DE once the budget grows. For more details on A1--A7, see
Section~\ref{ass}.}
\label{fig:tree}
\end{figure}
\subsection{Why we Compare Assumptions, not Tools?}
\label{assumptools}
Most SE optimizer studies end in a leaderboard
which states that some tool $A$ beat tools $B$ and $C$ on a
handful of benchmarks, so $A$ is reported the winner. But a leaderboard entry is a
fact about a \emph{tool} under one setup. It does not say \emph{why} the tool won, and
the ``why'' is the only part that carries to a new task. Two problems make the
leaderboard, on its own, a weak basis for choosing an optimizer.

First, the comparison is often unfair: a new method is matched against recent rivals,
not the simplest baseline. Only one in twenty LLM-for-SE papers used a simpler
approach~\cite{hou2024large}, and a plain sampler has matched far heavier search
whenever it was finally tested~\cite{chen2019sampling,nair2018finding,Amiraliminimaldata}.
With the baseline absent, a win is unreadable: the method may have won because it
suited the problem, or merely because its opponents were weak. Second, and more
fundamental, even a \emph{fair} win is condition-specific: across our 106 tasks no
family of optimizer leads on more than \textbf{37}, and the leader at $B{=}30$ is
usually not the leader at $B{=}200$ (\S\ref{sec:rq2}).  

What \emph{does} transfer is the reason behind the win. Every optimizer rests on a
\textbf{core assumption about the domain};
e.g. as studies in this paper,
assumptions   about the shape of the
\emph{fitness landscape}, the surface mapping each configuration to its measured
quality. Hill climbing, for instance, only makes sense if that surface is locally
smooth, so small steps yield small, informative changes~\cite{russell2009artificial}.
Other methods bet on decomposability, on a cheap surrogate standing in for evaluation,
and so on (Section~\ref{sec:related} reads one such assumption out of each optimizer's
originating paper). A benchmark tests the \emph{tool}, but what a practitioner needs to
know is whether the \emph{assumption} holds for their task and budget.

This is why a study that merely ranked $N$ algorithms would be incomplete: it would
name a winner without explaining what about the domain made it win, and so could not
guide anyone facing a different domain. We therefore organize the experiment around
the binary tree
of Figure~\ref{fig:tree}.
Each branch of the tree is a separate experiment with    two optimizers that differ in
essentially one assumption. The results
from these experiments then reads as evidence about \emph{which assumption
is useful, and under which budget}
(knowledge that carries to a new task) rather than as
one more brand ranking. 

Figure~\ref{fig:tree} summarizes domain assumptions 
using the  notation described in Section~\ref{ass}.  The green
check marks show winner results represented later in this paper,

\begin{finding}
\textbf{From tools to assumptions.} A leaderboard says which tool won, not why. So, the conclusion may not
transfer.  We compare \emph{assumptions} instead: each match isolates one, across
budgets, against a random floor. The by-product is a cheap, budget-aware selection
guide.
\end{finding}

\subsection{Why  Evaluate with 
Constrained Budgets?} 
Another special feature of this
evaluation is that we constrain evaluations budgets to just a few dozen. Why?

In SE, enumerating candidate choices is often cheap; but understanding their consequences is not. A single evaluation can be very expensive  e.g. scoring a build can mean a recompile
and a full test-suite run, and labeling one row of a project-health table can cost an
expert an hour or more~\cite{Amiraliminimaldata}.
It is possible to 
{\bf ask human experts for labels}  but that process is slow and often  error-prone~\cite{easterby1980design}.
For example, labeling just one row of a project-health table can cost an expert an hour or more, so even a few dozen  of cases can take up a week~\cite{lustosa2024learning,valerdi2010heuristics}.
Some researchers
{\bf mine historical logs} to deliver labels in bulk.
The quality of such labels is highly variable. One study found that 90\% of supposed technical-debt “false positives” were themselves mislabeled~\cite{yu2020identifying}.
The  same pattern recurs across security~\cite{wu2021data}, static analysis~\cite{kang2022detecting}, and defect datasets~\cite{shepperd2013data}.
As to using other methods to auto-label examples:
  regex-based heuristics can
  be very cheap but coarse-grained \cite{kamei2012large}. LLMs can assist but cannot serve as the final word~\cite{ahmed2025llmsreplacemanualannotation}.

This labeling cost is why it is prudent to score optimizers by the number of evaluations
they require.  It also explains why
a realistic evaluation budget may be 
dozens, not the thousands a generic hyperparameter optimization benchmark
assumes~\cite{eggensperger2013}. 

We note that the budget $B$  is not a constant to fix and forget:
Figure~\ref{fig:stakes} (right panel) puts numbers on the catch—the single best optimizer
differs between $B{=}30$ and $B{=}200$ on \textbf{50\%} of tasks, and \textbf{58\%} of
tasks change their best optimizer at least once across budgets. A comparison that
fixes one budget therefore risks conclusions that do not hold at other budgets.

\begin{finding}
\textbf{Budget is a first-class variable.} Evaluations cost CPU-years and expert
hours. Engineer may be able to  afford only dozens of evaluations, and which optimizer wins \emph{moves}
as that budget increases. An honest selection guide must be indexed by the budget, not
stated for a single one.
\end{finding}

\section{Related Work and Optimizer Selection}
\label{sec:related}

The no-free-lunch theorem makes exhaustive comparison pointless~\cite{wolpert1997no}: rather than
enumerate the vast SBSE and hyperparameter-optimization
menu~\cite{Amiraliminimaldata,harman2012search}, we categorize it and sample one
representative per category.

Reading the
originating paper of each optimizer family, we asked one question: \emph{what must be
true of the fitness landscape for this method to make sense?}. This clustered the
answers into seven assumption families (Table~\ref{tab:assumptions}), each 
of which is a testable
bet paired with the limitation that breaks it (and an SE example where it matters). From each family we take representatives that are \textbf{canonical} for some assumption (a basic version, not a tuned hybrid, so a win is attributable to the assumption), have a clear \textbf{specification} that we follow as published (a public implementation where one exists, otherwise the originating paper's algorithm implemented faithfully), and carry \textbf{recent SE relevance}. This yields the \textbf{20 black-box optimizers} of Table~\ref{tab:inventory},
each mapped to its assumption, role, origin, and SE uses. These cover six of the seven
assumptions; the seventh (A6, reinforcement learning) is described below for completeness
but excluded from the tournament, since it needs thousands of environment steps and so
cannot operate at the few-dozen-label budgets we study. 

We leave to future work methods that require resources our setting cannot provide: high-dimensional
Bayesian~\cite{cowen2022hebo,eriksson2019scalable}, causal~\cite{chen2026promisetune,iqbal2022unicorn},
knob~\cite{kanellis2022llamatune,van2017automatic,zhang2019end},
reinforcement-learning~\cite{li2019qtune}, 
LLM-based~\cite{romera2024mathematical,ye2024reevo}
optimizers.

\begin{table}[tbp]
\centering
\caption{Landscape assumptions behind SBSE optimizers: read from each method's
originating paper, with the limitation that breaks each and an SE task where it matters.}
\label{tab:assumptions}
\footnotesize
\setlength{\tabcolsep}{4pt}
\renewcommand{\arraystretch}{1.12}
\begin{adjustbox}{max width=\columnwidth}
\begin{tabular}{@{}clp{4.2cm}@{}}
\toprule
\textbf{ID} & \textbf{Assumption} & \textbf{Limitation \& SE example}\\
\midrule
A1 & Local continuity & Fails on rugged landscapes. \emph{Compiler-flag tuning \cite{chen2023compiler}.}\\
A2 & Memory / diversification & Fails on deceptive neighborhoods. \emph{Test-suite minimization \cite{diaz2008tabu}.}\\
A3 & Building-block decomposability & Crossover destroys coupled structure. \emph{Program repair \cite{legoues2011genprog}.}\\
A4 & Pareto incomparability & Wasteful under clear preferences. \emph{Software remodularization  \cite{mkaouer2015many}.}\\
A5 & Surrogate feasible & Surrogate assumptions fail under few labels. \emph{DBMS Knob tuning \cite{kanellis2022llamatune}.}\\
A6 & Sequential reward signal & Fails under heavy delay. \emph{Cloud DB tuning \cite{zhang2019end}.}\\
A7 & Low intrinsic dimensionality & Fails when data fills the space. \emph{High-dim.\ feature models \cite{sayyad2013value}.}\\
\bottomrule
\end{tabular}
\end{adjustbox}
\vspace{-0.3cm}
\end{table}

\subsection{ Related work and Assumption Families}\label{ass}
\textbf{Trajectory and local search (A1, A2).} These methods keep a \emph{single}
candidate and repeatedly replace it with a better neighbor, in effect walking step by
step across the landscape rather than maintaining a population. {\em Hill climbing} is
the textbook embodiment of the continuity bet: it always steps to the best nearby
point and halts when none improves~\cite{russell2009artificial}. {\em Simulated
annealing} adds a cooling schedule that occasionally accepts a \emph{worse} move (more
readily early on) so the search can escape local optima~\cite{kirkpatrick1983optimization}.
It has been used for search-based refactoring~\cite{mkaouer2015many}. The \emph{(1+1) Evolution Strategy} keeps the same single-incumbent form but replaces the fixed neighborhood with a self-adaptive Gaussian step that tunes its own size (the 1/5 rule) \cite{rechenberg1973evolutionsstrategie}, and has been used for test-data generation \cite{arcuri2018test}. A diversification
family answers ruggedness by avoiding places already seen: {\em tabu search} forbids
revisiting recent points, with an aspiration rule that lifts the ban for an exceptional
gain~\cite{glover1989tabu}, and appears in configuration transfer and structural
testing~\cite{lin2015tca,diaz2008tabu}. {\em Iterated local search} kicks the incumbent
into a new region and re-optimizes~\cite{lourenco2003iterated,sabar2018genetic}. These methods are cheap per step and warm
up fast.

\textbf{Population and evolutionary search (A3).} These methods keep a
\emph{population} of candidates and breed new ones from the fittest, so good partial
solutions can spread and combine. \emph{Genetic algorithms} recombine two parents under
the building-block bet that good solutions are assembled from good
parts~\cite{holland1975adaptation}, extended to programs by \emph{genetic programming}
and automated repair~\cite{koza1992genetic,legoues2011genprog}. \emph{Differential evolution} forms a new
candidate by adding the scaled difference of two population members to a third, keeping
the child only if it beats its parent~\cite{storn1997differential}. \emph{Particle
swarm} nudges each candidate along its own best and the swarm's best
directions~\cite{kennedy1995particle}; \emph{estimation-of-distribution} algorithms
replace recombination with a probability model fitted to the best-so-far and then
sample from it~\cite{muhlenbein1996recombination}.

\textbf{Multi-objective and Pareto search (A4).} When objectives conflict there is no
single best configuration, only \emph{trade-offs}: the \emph{Pareto frontier} are  solutions where no objective can be improved without worsening another. These
methods evolve a population to approximate that frontier.
\emph{NSGA-II} sorts candidates into nondomination layers and breaks ties by crowding,
favouring sparser regions so the frontier stays diverse~\cite{deb2002fast}.
\emph{SPEA2} keeps an external archive in which each solution is scored by how many
others it dominates~\cite{zitzler2001spea2}. \emph{MOEA/D} splits the problem into many
scalar subproblems and solves them together, each pulling toward a different part of
the frontier~\cite{zhang2007moead}, while \emph{reference-point} methods use fixed
directions to scale to many objectives~\cite{deb2014evolutionary}. SBSE has long recommended full Pareto search over weighted or aggregated methods for multi-objective problems~\cite{chen2023weights}. This has only recently been questioned. For instance, meta-multi-objectivization, can reshape the problem so that fewer objectives suffice~\cite{chen2024meta}.

\textbf{Surrogate and model-based search (A5).} These methods fit a cheap
\emph{surrogate} model of the response from the points labelled so far, then use it to
decide what to evaluate next, spending real evaluations only where the model expects a
payoff. \emph{Bayesian optimization} fits a probabilistic surrogate and picks the next
point with an acquisition function that trades expected gain against
uncertainty~\cite{snoek2012practical}. \emph{SMAC} uses a Random Forest surrogate,
letting it handle the categorical and non-continuous spaces that Gaussian processes
handle poorly~\cite{hutter2011sequential}. \emph{TPE} instead models the densities of
good and of ordinary configurations and samples where the ratio favors good~\cite{bergstra2011algorithms}. The configuration tuning community has produced
many task-specific tuners: knob tuners~\cite{kanellis2022llamatune,van2017automatic},
sequential model-based configuration~\cite{nair2018finding}, causal approaches~\cite{iqbal2022unicorn,chen2026promisetune}, configuration-space reduction~\cite{cao2024etune}, drift adaptation~\cite{xiang2026dually}, co-evolutionary tuning~\cite{xiong2025cotune}, and Bayesian compiler
autotuning~\cite{chen2021efficient,zhu2023compiler}. Crucially, recent work shows surrogate's \emph{predictive accuracy} is a poor proxy for quality of the configurations a method ultimately selects~\cite{chen2025accuracy,chen2025surrogatefaces}, which motivates judging each optimizer by the configurations it delivers rather than by any internal model's loss.

\textbf{Reinforcement learning (A6).} Reinforcement learning learns a \emph{policy} by
trial and error from a reward that may arrive only after a sequence of
actions~\cite{watkins1989learning,mnih2015human}. In database
tuning this yields end-to-end tuners that treat one full benchmark run as a single
environment step and its measured quality as the reward~\cite{li2019qtune,zhang2019end}.

\textbf{Data-light and geometric search (A7).} These methods bet that the data has low
\emph{intrinsic dimensionality} (i.e., it collapses into a few regions), so a few dozen
well-chosen labels capture most of its structure, and often they read the geometry of sampled
points rather than fit a model~\cite{Amiraliminimaldata}.  \emph{LINE} picks
centroids covering high-variance regions~\cite{ganguly2026lowgodatalightse}. \emph{EZR} is a
distance-based active learner that labels points near the boundary between the current
best and the rest~\cite{Amiraliminimaldata, ganguly2026lowgodatalightse}. \emph{DODGE} discards
configurations that fall in the same $\epsilon$-bin and excels when intrinsic
dimensionality is low~\cite{agrawal2019dodge}. \emph{SWAY} recursively bisects the space
by distance to sample a few representatives~\cite{chen2019sampling}. The baseline, \emph{Random probing} labels
a random budget and returns the best~\cite{bergstra2012random}, and can often yield surprisingly strong results~\cite{bergstra2012random}.

\textbf{Algorithm selection and instance-space analysis.} The dominant alternative to
running a tournament is to \emph{predict} the winner from cheap problem features.
\emph{Instance-space analysis} projects each problem into a low-dimensional space of
landscape features and draws ``footprints'' marking where each algorithm
wins~\cite{smith2023instance}; it has been applied to fitness landscapes
generally~\cite{albunian2020causes,aleti2017analysing} and to search-based test
generation specifically~\cite{neelofar2023instance}. We use the same feature family in
RQ4, but we \emph{test} the predictive claim rather than assume it, and we add the
variable this literature omits: the evaluation budget.

\textbf{What we exclude, and why.} No study can explore all algorithms since there are so many of them.   We elect to exclude causal tuners such as
PromiseTune~\cite{chen2026promisetune} and Unicorn~\cite{iqbal2022unicorn}: these are hybrids that build a causal model on top of their search, whereas our inventory takes basic optimizers whose result traces to a single assumption. We also exclude multi-fidelity
methods such as BOHB~\cite{falkner2018bohb} and DEHB~\cite{awad2021dehb}, which need
a fidelity ladder the tabular tasks do not provide.
Further, we do not study high-dimensional Bayesian
methods such as TuRBO~\cite{eriksson2019scalable}, whose machinery targets
hundred-plus-dimensional spaces far larger than ours.
Nor do we study LLM-driven
optimizers~\cite{zhang2025using,meyerson2024language,romera2024mathematical}, since these have such a large computational cost it would be prohibitively expensive (to say the least)
to run this study on those kinds of algorithms.  
\section{Method:    Assumption-Indexed Tournaments}
\label{sec:method}


\subsection{Evaluating a configuration}
\label{sec:surrogate}

MOOT tasks are represented as lookup tables, not models.
To evaluate new solutions,
the simplest approach would be to use dependent values from the nearest neighbor of the proposed new solution. While
this evaluation approach has some precedence in the literature
 (see Pfisterer and Zela et al.~\cite{pfisterer2022yahpo, zela2022}),
 it tends to reward solutions that are close
 to existing solutions. Hence, it may not
 be a fair oracle from
 membership-query methods like SMAC  that propose points
\emph{between} rows. 

An alternate approach,
taken by Eggensperger and Hutter and Nair et al.~\cite{eggensperger2015efficient, eggensperger2018efficient, hutter2011sequential,nair2018finding} is to build a random forest ensemble using all the training data.  This is useful since SE problems contain categorical and numeric variables, which random forests can handle.
Also, SE problems are   rarely smooth  since every ``if'' statement in a program can divide the internal state space into yet another region with different properties.
Random forests are good
for   multi-branching and building one model per branch.
For these reasons, this study uses random forests as the solution oracle. 

As to other details: to avoid a tuning-the-tuner conflation, we run every
method at its originating-paper defaults (Table~\ref{tab:hyper}). We sweep four
evaluation budgets, $B\in\{30, 50, 100, 200\}$ measured configurations, and
repeat every (method, task, budget) cell over 20 random seeds; all reported
statistics are over those repeats. The full design is thus 20
optimizers $\times$ 106 tasks $\times$ 4 budgets $\times$ 20 seeds, which, with
the multi-objective tree and the all-pairs ranking used as a check, comes to the
order of $1.8\times10^5$ search runs. Each run also pays for fitting and
querying the per-task surrogate. Aggregated over the per-run wall-clock times in
our logs (summarized in Figure~\ref{fig:runtime}), the study consumed roughly
\textbf{14{,}000 CPU hours}.

(Aside: We mention this CPU cost because it  is itself a finding. It is why practitioners do not run this
tournament themselves, and therefore why the \emph{cheap-to-obtain} guide  (at the end of this paper) that can cheaply reproduce our verdict is
of much practical value.)

\begin{table}[tbp]
\centering
\caption{Hyperparameters, set to each method's originating-paper defaults to avoid a
tuning-the-tuner confound.}
\label{tab:hyper}
\scriptsize
\setlength{\tabcolsep}{4pt}
\renewcommand{\arraystretch}{1.1}
\begin{tabular}{@{}ll@{}}
\toprule
\textbf{Optimizer} & \textbf{Key settings (defaults)}\\
\midrule
Hill Climbing       & greedy step; full neighbourhood\\
Simulated Annealing & geometric cooling $0.95$; random neighbour\\
Iterated Local Sea. & perturbation on stagnation; accept-better\\
Tabu Search         & tabu tenure $7$\\
Genetic Algorithm   & pop.\ $10$; uniform crossover $0.9$; mut.\ $1/n$\\
(1+1)-ES            & single parent; Gaussian mut.; $1/5$ rule\\
EDA                 & pop.\ $10$; univariate model; top-$50\%$\\
PSO                 & pop.\ $10$; $w{=}0.7$, $c_1{=}c_2{=}1.49$\\
SMAC                & RF surrogate; EI acquisition; init $10$\\
DE                  & pop.\ $10$; $F{=}0.8$, $CR{=}0.9$; rand/1/bin\\
TPE                 & $\gamma{=}0.25$; startup $10$\\
LINE (kpp)          & $k$ from budget; centroid init\\
SWAY                & recursive median-distance bisection\\
EZR                 & distance acquisition; init labels $4$\\
DODGE               & $\epsilon{=}0.2$; $\epsilon$-bin tabu\\
Random Search       & uniform sampling (floor)\\
NSGA-II             & pop.\ $10$; SBX; polynomial mutation\\
SPEA2               & pop.\ $10$; archive $10$\\
SMS-EMOA            & pop.\ $10$; hypervolume contribution\\
MOEA/D              & pop.\ $10$; $5$ neighbours; Tchebycheff\\
\bottomrule
\end{tabular}
\end{table}

\subsection{Tasks and scoring}
\label{sec:scoring}
Each task is a set of configurations with one or more measured objectives, from the
SE datasets of MOOT (Table~\ref{tab:datasets}). We orient and normalize every objective to
$[0,1]$ so that $0$ is best, and score a configuration by its \emph{distance to
heaven}: the normalized Euclidean distance to the ideal point where every objective is
at its best,
\begin{equation}\label{dtoh}
\dtoh(c) \;=\; \sqrt{\frac{1}{|O|}\sum_{o\in O} \big(o(c)\big)^2}\,,
\qquad o(c)\in[0,1],\ 0=\text{best},
\end{equation}
where $O$ is the objective set and $o(c)$ the oriented, normalized value of objective
$o$. Lower is better and the ideal point scores $0$. We use \dtoh{} because it
collapses single- and many-objective quality onto one axis, letting the same
tournament rank a hill climber against an NSGA-II run. It rewards configurations
\emph{balanced} near the ideal rather than extreme on one axis. It follows the data-light SE line we build on~\cite{Amiraliminimaldata,lustosa2024learning,nair2018finding}.

\textbf{Why best-point, not the frontier.} For multi-objective tasks a method may
return a set of nondominated configurations, but a team ships \emph{one}. We
therefore score each method by the single returned configuration with the lowest
\dtoh{}, the point a practitioner would actually deploy. This reduction could in
principle flatter scalar methods, so we do not take it on faith: in RQ3 we re-run
the multi-objective comparison with frontier-aware metrics (IGD, GD, hypervolume,
Section~\ref{sec:rig}). The single-objective methods still lead: at equal budget EZR/ DE win on IGD, GD, and hypervolume too. So the best-point reduction is not what drives the single-versus-multi result.

\section{Experimental Rig}
\label{sec:rig}
 \begin{figure*}[tbp] \centering 
\includegraphics[width=0.6\textwidth]{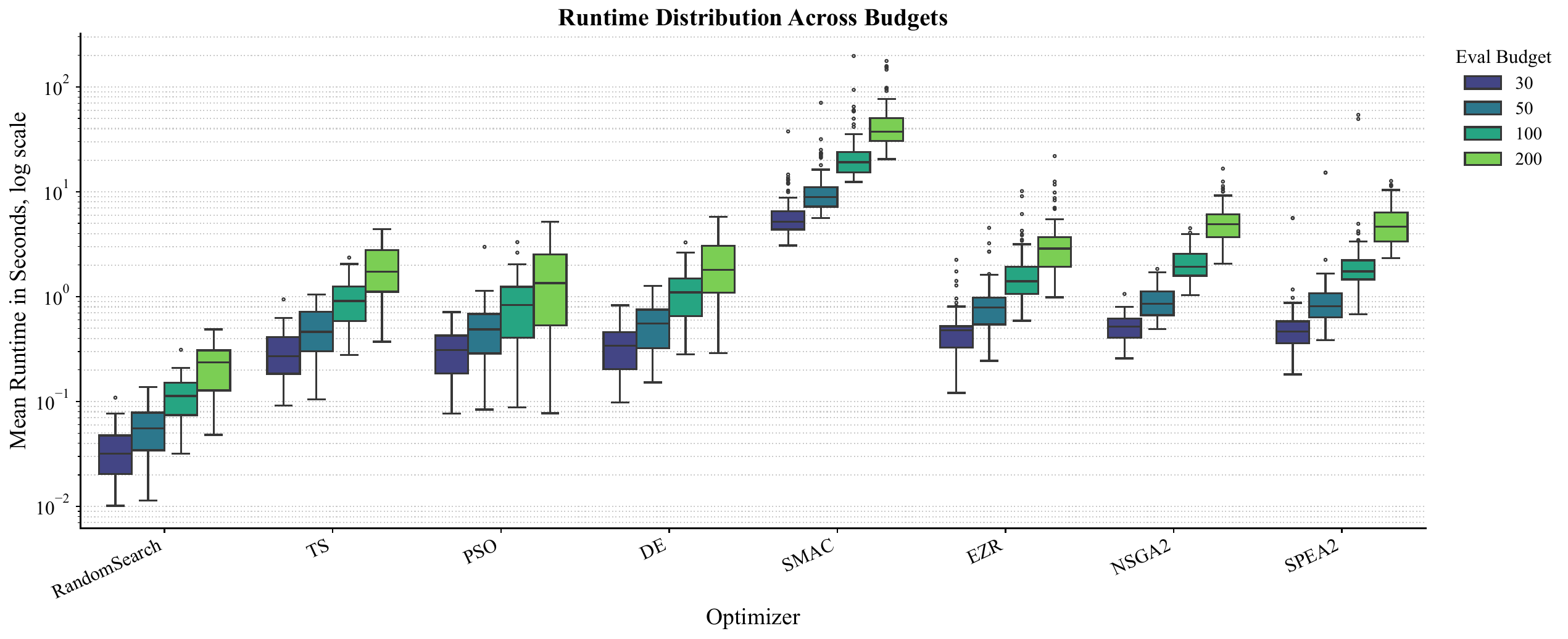} 
\caption{\textbf{Search cost is real and it is the reason a free guide matters.} Per-run wall-clock distributions for every wining optimizer at each budget, aggregated to the $\approx$\textbf{14{,}000 CPU hours} the full study consumed. } 
\label{fig:runtime} \end{figure*}

\textbf{Regret reduction metric.} Raw \dtoh{} values are not comparable
across tasks of different difficulty, so in addition to \dtoh, we report how much of the available regret
a method removes. For a task, let $\dtoh_{\mathrm{rand}}$ be the expected quality of
random search at the same budget and $\dtoh^\star$ the best achievable quality on
the task. A method achieving $\dtoh_m$ removes the fraction
\begin{equation}
\rho \;=\; \frac{\dtoh_{\mathrm{rand}} - \dtoh_m}{\dtoh_{\mathrm{rand}} - \dtoh^\star}\,,
\label{eq:regret}
\end{equation}
of the random-to-optimal gap, with $\rho=1$ matching the best configuration on the
task and $\rho=0$ matching blind sampling. We report the median $\rho$ over tasks
at each budget. This framing keeps the random floor visible at all times and makes
``earned the right to beat random'' measurable.

\textbf{Multi-objective sanity metrics.} To check that the best-point reduction of
Section~\ref{sec:scoring} does not distort the single-versus-many comparison, RQ3
additionally scores the multi-objective methods with three frontier-aware measures:
generational distance (GD), the mean distance from the returned set to the reference
frontier constructed from the best-known points pooled over all optimizers' results, inverted generational distance (IGD), the mean distance from the reference
frontier to the returned set, which rewards coverage, and hypervolume (HV), the
volume of objective space dominated by the returned set relative to the best-known reference set pooled over all optimizers' results~\cite{zitzler2001spea2,beume2007sms,deb2014evolutionary}. Lower GD
and IGD and higher HV are better. These are reported only as a cross-check; the
tournament itself is decided on \dtoh{}.


\textbf{Statistics.} Every match and every ranking uses the same nonparametric
procedure. Over the 20 seeds we apply Scott-Knott clustering~\cite{scott1974cluster}, which recursively
bisects the methods into ranked groups and accepts a split only when a bootstrap
test finds the groups distinct \emph{and} the Cliff's delta effect size between
them is non-negligible. Methods that
land in the same group are statistically indistinguishable and share a rank.

\section{Results}
\label{sec:results}

We organize our results around the following questions:

\begin{itemize}

\item RQ1: Is optimization worth doing?

\item RQ2: Does the budget change which optimizer wins?

\item RQ3:  Do these optimizers really differ?

\item RQ4: Can we cheaply predict the winner?

\end{itemize}
Of these, RQ4 is of the most practical importance.
Figure~\ref{fig:runtime} shows the runtimes for a sample of the optimizers used in this study. Note that
the  runtimes increase exponentially with the labeling budget.
Exponential evaluation costs mean that most
practitioners cannot afford to repeat this paper's study.
They require some way to quickly peek at a problem, then decide what optimizer to use.

\subsection{RQ1: Is   optimization useful?}
\label{sec:rq1}
As argued in Figure~\ref{fig:stakes}: even a few dozen labels are enough to find large optimizations. 
Hence, we report from Figure~\ref{fig:stakes}:

\begin{finding}
\textbf{RQ1.} 
The payoff from optimization are real,
even with very few labels. 
At budgets of just  $B=30$ and $B=200$,
our best optimizers improve optimization from \textbf{30.7\%} (at $B{=}30$) to \textbf{74.5\%}
(at $B{=}200$) of the random-to-optimal regret.
\end{finding}
 
\subsection{RQ2: Does budget change which optimizer wins?}
\label{sec:rq2}
 Table~\ref{tab:tournament} applies our statistical methods at each level of the trees in Figure~\ref{fig:tree}. As can be seen,  as the budget increases, the ``best'' optimizer changes:

\begin{table}[!b]
\centering
\caption{Tournament winners by stage and budget (Scott-Knott, 20 seeds). The
single-objective champion is EZR at tight budgets, DE once the budget grows.
Bottom rows: median regret reduction $\rho$ and champion mean \dtoh{}.}
\label{tab:tournament}
\footnotesize
\setlength{\tabcolsep}{5pt}
\renewcommand{\arraystretch}{1.12}
\begin{tabular}{@{}l cccc@{}}
\toprule
& \multicolumn{4}{c}{\textbf{Evaluation budget $B$}}\\
\cmidrule(l){2-5}
\textbf{Bracket stage} & \textbf{30} & \textbf{50} & \textbf{100} & \textbf{200}\\
\midrule
SF-A \;(Trajectory)      & TS      & TS      & TS      & TS\\
SF-B \;(Population)      & PSO     & PSO     & DE     & DE\\
SF-C \;(Models)         & SMAC    & SMAC    & SMAC      & SMAC\\
SF-D \;(Samplers)       & EZR     & EZR     & EZR     & EZR\\
\addlinespace[1pt]
SF-AB (Local)           & PSO     & PSO     & DE     & DE\\
SF-CD (Representation)  & EZR     & EZR     & SMAC      & SMAC\\
\addlinespace[1pt]
T1-SO champion          & EZR     & EZR     & DE      & DE\\
T2-MO champion          & NSGA2   & NSGA2   & SPEA2   & SPEA2\\
\midrule
\textbf{Grand Final}    & \textbf{EZR} & \textbf{EZR} & \textbf{DE} & \textbf{DE}\\
\midrule
Median $\rho$ (\%)      & 30.7 & 42.3 & 52.0 & 74.5\\
Champion mean \dtoh{}   & 0.21 & 0.19 & 0.17 & 0.16\\
\bottomrule
\end{tabular}
\vspace{-0.3cm}
\end{table}


\begin{finding}
\textbf{RQ2.} The budget chooses the optimizer. When data is scarce,  $B\le 50$  \textbf{EZR} works best. But at
larger budgets $B\ge 100$ \textbf{DE} works best. 
\end{finding}

\subsection{RQ3: Do these optimizers really differ?}
\label{sec:rq3}
\begin{table}[!t]
\centering
\caption{Single- vs.\ multi-objective search on 81 multi-objective tasks
(best-point \dtoh{}, IGD, GD, HV; median over tasks). At every matched budget the
single-objective methods lead or tie; multi-objective search needs $5\times$ the budget
to reach EZR@200, and even then EZR@1000 stays ahead.}
\label{tab:somo}
\footnotesize
\setlength{\tabcolsep}{5pt}
\renewcommand{\arraystretch}{1.12}
\begin{tabular}{@{}l c c cccc@{}}
\toprule
\textbf{Method} & objectives & \textbf{$B$} & \textbf{best-pt \dtoh{}}$\downarrow$
& \textbf{IGD}$\downarrow$ & \textbf{GD}$\downarrow$ & \textbf{HV}$\uparrow$\\
\midrule
\multicolumn{7}{@{}l}{\emph{$B{=}200$}}\\
EZR   & single & 200 & 0.18 & \textbf{0.12} & 0.05 & \textbf{0.97}\\
DE    & single & 200 & \textbf{0.16} & 0.13 & \textbf{0.03} & 0.93\\
NSGA-II   & multi & 200 & 0.22 & 0.16 & 0.08 & 0.91\\
SPEA2     & multi & 200 & 0.29 & 0.23 & 0.14 & 0.73\\
MOEA/D    & multi & 200 & 0.22 & 0.16 & 0.07 & 0.92\\
SMS-EMOA  & multi & 200 & 0.40 & 0.31 & 0.15 & 0.48\\
\midrule
\multicolumn{7}{@{}l}{\emph{$B{\approx}1000$ ($5\times$)}}\\
EZR   & single & 1000 & \textbf{0.15} & \textbf{0.10} & 0.03 & \textbf{1.01}\\
DE    & single & 1000 & \textbf{0.15} & 0.13 & \textbf{0.02} & 0.94\\
NSGA-II   & multi & 1000 & 0.18 & \textbf{0.10} & \textbf{0.02} & \textbf{1.01}\\
SPEA2     & multi & 1000 & 0.25 & 0.21 & 0.08 & 0.80\\
MOEA/D    & multi & 1000 & 0.18 & 0.11 & \textbf{0.02} & 0.98\\
SMS-EMOA  & multi & 1000 & 0.37 & 0.28 & 0.11 & 0.52\\
\bottomrule
\end{tabular}
\par\smallskip\smallskip
{\footnotesize At equal budget the single-objective methods win or tie every metric.
Multi-objective search needs about $5\times$ the budget to catch up for \dtoh{}.}
\par\smallskip
{\footnotesize Let $d$ be the best-known front (pooled over all optimizers) and $a$ one
optimizer's returned set. GD is the mean distance from each point in $a$ to its
nearest in $d$; IGD reverses this, from each point in $d$ to its nearest in $a$
(rewarding coverage); HV is the objective-space volume $a$ dominates relative to
a worst-case reference point. Lower GD and IGD are better (closer to the
best-known front); higher HV is better (more of the space dominated).}
\end{table}
 
A discussion of all the results across the  tree in Figure~\ref{fig:tree}
(explaining every branch, every win, every
tie) is its own paper. In this section, we report on three standout features.

\textbf{A few optimizers are significantly
better.} As reported above \ezr{} and \de{} do
stand out from the pack.

\textbf{But many optimizers   tie.} Across the branches of Figure~\ref{fig:tree}, the
great majority of head-to-head matches end in a Scott-Knott tie: at a given budget,
both contestants land in the same rank. These methods use  very different machinery,
yet on SE data, at budgets an engineer can afford, they mostly deliver the same results.
This is a match-level echo of the family-level result discussed in \S\ref{assumptools} (no family leads on more
than 37 of 106 tasks) and fits the No-Free-Lunch view: once the budget is tight and
the data is fixed, many of these methods are interchangeable, so the field is narrower than the literature's steady stream of new optimizers implies. A few methods still separate from the pack, and which one leads shifts with the budget, the structure our guide is built to capture.

\textbf{And a few of the losses are surprising; e.g.  single vs.\ multi-objective.} The SE
optimization literature holds that multi-objective  methods beat methods that
collapse many goals into one aggregate. To our knowledge this has not been checked
before on a large SE sample over such a tournament structure. Our setup lets us test it directly: we treat each multi-objective task as a single-objective one by aggregating its objectives into a single \dtoh{} score (Equation~\ref{dtoh}, as EZR does), run the single-objective methods on it, and then score the full set of configurations each method evaluated over its search trajectory with frontier-aware metrics (GD, IGD, HV), defined in Table~\ref{tab:somo}. This puts single- and multi-objective methods on the same frontier-based footing.Note that:
\begin{itemize}
\item  {\em lower} \dtoh, GD and IGD are {\em better}; \item  {\em higher} HV is
{\em better}.
\end{itemize}

Table~\ref{tab:somo} compares single- and multi-objective reasoners.
Best per column
are shown in  \textbf{bold}. On all of the measures, multi-objective methods cannot surpass single-objective ones. Even when using $5\times$ more budget, none of the multi-objective optimizers can surpass single-objective ones on equal budgets.  And the cost asymmetry is stark, Table~\ref{tab:somo} shows that NSGA-II needs 1000 samples to match what EZR reaches in 200.

\begin{finding}
\textbf{RQ3.} Yes, but less than the literature suggests. Most matches tie: at SE budgets, very different optimizers deliver the same result, and only a few separate themselves from the pack. Overall, EZR and DE lead but which optimizer is best depends on the task and budget and no single method wins everywhere. Even the single- vs.\ multi-objective divide collapses: at least one single-objective search matches or beats multi-objective methods at equal budget.
\end{finding}

\begin{figure}[!th]
\centering
\includegraphics[width=0.8\columnwidth]{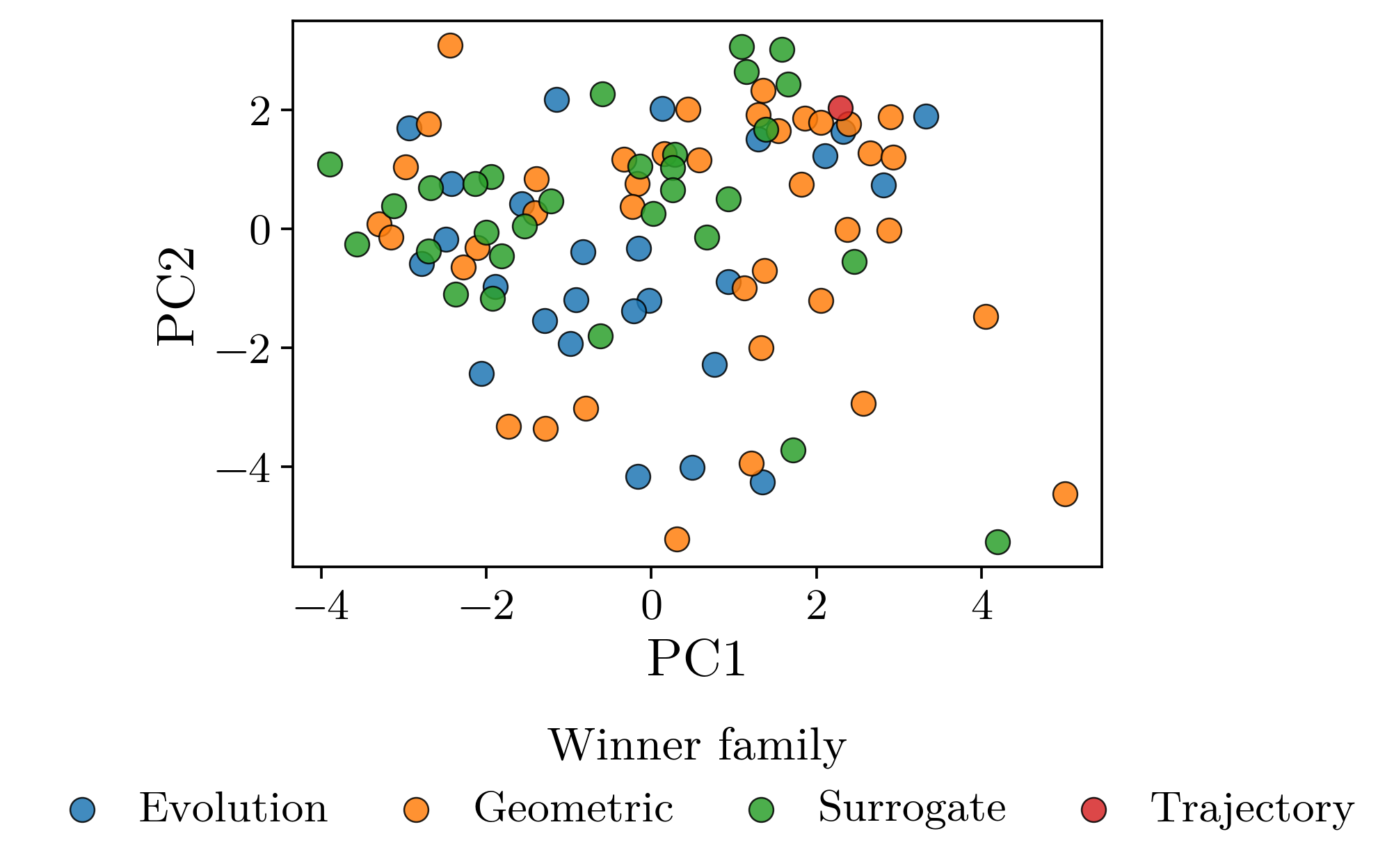}
\caption{\textbf{Static instance clustering features cam not find
the right optimizer} A two-dimensional projection of
the tasks in landscape-feature space, colored by the budget-200 winning family.
If cheap features predicted the winner, the colors would separate into clean
footprints. They do not: the regions overlap heavily, which is the visual companion
to the \textbf{44.2\%} accuracy.}
\label{fig:instance}
\end{figure}

\subsection{RQ4:  Can we predict the best optimizer?}
\label{sec:rq4}

In summary, so far we have said that prior work was incomplete on important context variables (the labeling budget) and misleading on others (the relative merits of single vs.\ multiple objective optimization). If prior advice cannot be trusted, then practitioners face a new domain with no reliable way to pick an optimizer. Hence we must ask, based on this study:
\begin{quote}
{\em What guidance can be offered for selecting optimizers for new domains?}
\end{quote}
This section explores  this question in two parts. First we report on a failed experiment with building a guidance tool based on instance clustering metrics.  
Secondly, we report a second experiment with building a guidance tool   that is far more effective (at predicting the right optimizer)
and uses domain features that are very cheap to collect.


\subsubsection{Guidance via instance clustering metrics (does not work)}
Prior work ~\cite{mersmann2011exploratory, smith2023instance,neelofar2023instance, kerschke2015detecting}
suggests  optimization difficulty can be modeled as a function of measurable properties of the problem instance or sampled search surface. Following that advice,  we collected
 metrics describing the relationship between the configuration values and the objective scores
(fitness-distance correlation \cite{jones1995fitness}, dispersion \cite{lunacek2006dispersion}, nearest-better clustering \cite{kerschke2019automated}, meta-model
$R^2$, skewness, kurtosis, and PCA-based geometry features \cite{mersmann2011exploratory, kerschke2015detecting}) from a
random sample of 100 rows of a  data set.
This was used to train a feature-based predictor (using Random Forests) \cite{smith2023instance, kerschke2019automated} of what family of optimizers
was most useful for a particular MOOT data
set (for a list of those family names,
recall Figure~\ref{fig:tree}). Since the family distribution was imbalanced in our results, we used a 10 times repeated stratified 10-fold cross validation \cite{kohavi1995study}.

The resulting classifier  only 
reached \textbf{44.2\%} accuracy for predicting what family of optimizer works best for that data set (for knowledge of what optimizer works best, we used our RQ2 results). As a stricter check, we also trained the same feature-based model to predict the exact optimizer, which reached only \textbf{34\%} accuracy. Thus, even when the feature-based approach is given the easier family-level target, it does not provide reliable guidance in our setting. 
  
Figure~\ref{fig:instance} explains why instance clustering metrics worked so poorly:  projected into feature space our
tasks do not separate into clean per-winner footprints (observe the region overlap).  Hence, we need to look beyond instance clustering metrics.



\subsubsection{Guidance via heatmap (works better)}
Given the failure of instance clustering metrics, we tried other
approaches.  
Since the goal was an easy-to-used guide,
we asked ``what  attributes could an  engineer   read off
a MOOT data  table at no cost, with no inference''.

We used two such task attributes, plus the labeling budget.
\begin{itemize}
\item
Whether the task has
conflicting objectives (optimizing  one thing hurts another goal;
e.g. developing cheaper software with fewer bugs);
\item
The shape of its input space: binary/SAT (i.e. the columns values are true,false), small-numeric, or
large-numeric.
\end{itemize}

Each task is assigned to one of six heatmap cells using two inexpensive structural attributes computed directly from its input table. The first distinguishes \textit{single-objective} tasks (one optimization objective) from \textit{multi-objective} tasks (two or more objectives). The second characterizes the input space. Let $|X_i|$ denote the number of distinct values of decision variable $i$. Since MOOT does not contain the real-world range and constraints of the decision variables, we approximate $|X_i|$ with the number of distinct decision variable values observed in MOOT. Hence, we estimate the search-space size as: $\sum_i \log_2(|X_i|)$, which is the base-2 logarithm of the Cartesian product of all decision-variable domains. 

A task is classified as:
\begin{itemize}
    \item \textbf{Binary/SAT}: at least 80\% of decision variables are binary ($|X_i|=2$);
    \item \textbf{Large-numeric}: not Binary/SAT and $\sum_i \log_2(|X_i|)\ge40$;
    \item \textbf{Small-numeric}: all remaining tasks.
\end{itemize}

The boundary $40$ is empirically defined. As a sensitivity check, we repeated the guide evaluation with $10,20,30,40,50 \text{, and}$ 60. The general accuracy remained similar for $10-40$ and dropped minimally ($\approx 6\%$ on average) only for higher values.

\begin{figure}[tbp]
\centering
\includegraphics[width=0.7\columnwidth]{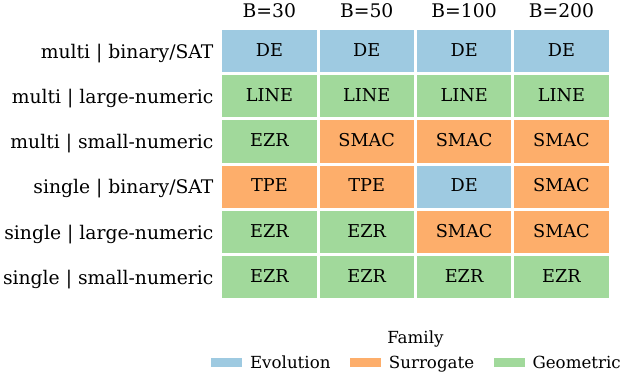}
\caption{\textbf{The free, budget-aware selection guide (RQ4).} Recommended
optimizer per objective structure, input-space shape, and budget, read off a task
table at zero cost. EZR wins single-objective, small-numeric cells at tight
budgets; DE and SMAC take over as input space or budget grows. Ties or beats a
hindsight oracle on \textbf{74.2\%} of held-out tasks (Table~\ref{tab:guideval}).}
\label{fig:guide}
\vspace{-0.4cm}
\end{figure}

Figure~\ref{fig:guide}
shows the resulting guide, a recommended optimizer per attribute cell per budget.
EZR owns the single-objective small-numeric cells at tight budgets; DE and SMAC take
over as the input space grows or the budget increases, exactly the migration RQ2
exposed.

The decisive test is whether this cheap-to-obtain guide \emph{generalizes}. We evaluate it
under 10 repeated shuffled 10-fold cross-validation stratified over aforementioned task classes. In each fold, we run the tournament (recall Fig. \ref{fig:tree} and produce a guide similar to Figure \ref{fig:guide}. For each budget levels, The guide is then applied to the held-out tasks in that fold and compared against a \emph{hindsight oracle}, 
defined as the tournament champion computed on those held out tasks at the same budget. 

The oracle is a deliberately hard baseline: it has seen
the test tasks. Table~\ref{tab:guideval} reports the result. The cheap-to-obtain guide ties or
beats this hindsight oracle on \textbf{74.2\%} of held-out tasks overall with zero probes. The instance features requiring 100 evaluations predicts the winning family only $~44.2\%$ of the time and even lower in terms of exact prediction, while this no-additional-probe guide, combined with the budget, recover the tournament winner about three times in four. 

\begin{finding}
\textbf{RQ4.} The best optimizer for a new SE task can be predicted cheaply. A guide keyed on objective conflict, input-space shape, and budget ties or beats a hindsight oracle on 74.2\% of held-out tasks, at zero probe cost. Instance clustering metrics failed to do so. A practitioner facing a new SE task needs only to read two attributes from their data table and consult Figure \ref{fig:guide}.
\end{finding}

\begin{table}[tbp]
\centering
\caption{Win rate of the budget-aware
guide under 10-seed stratified 10-fold cross-validation (200 train/test partitions).}
\label{tab:guideval}
\footnotesize
\setlength{\tabcolsep}{8pt}
\renewcommand{\arraystretch}{1.12}
\begin{tabular}{@{}l c@{}}
\toprule
\textbf{Evaluation budget $B$} & \textbf{Win rate vs.\ oracle (\%)}\\
\midrule
30 & $81.0 \pm 9.4$ \\
50 & $72.0 \pm 7.5$ \\
100 & $71.0 \pm 11.4$ \\
200 & $73.0 \pm 7.8$ \\
\midrule
\textbf{Overall} & $\mathbf{74.2 \pm 5.6}$\\
\bottomrule
\end{tabular}
\vspace{-0.4cm}
\end{table}


\subsection{Threats to Validity}
\label{sec:threats}
\textbf{Construct validity.} Our quality metric is \dtoh{}, and reducing a
multi-objective result to its best single point could in principle favor scalar
methods. This is why we further evaluate the multi-objective tasks with
frontier-aware metrics (IGD, GD, HV) RQ3.

A second
construct concern might be that the response surface is a fitted surrogate rather than
ground truth. we tuned the surrogate and used one tuned surrogate per task, held fixed
across all optimizers, so any bias is shared by every method and cannot change their
relative ranking, and we judge methods by delivered \dtoh{} rather than by surrogate
accuracy.


\textbf{External validity.} All tasks come from
MOOT and there might be other data sets for which these results
do not hold. This is a problem
of any empirical study since no study can exercise
all data sets. To mitigate this, all we can do is:
\begin{itemize}
\item
Make our case study space as large as possible.
The 106 tasks span
seven SE domains from configuration and performance tuning to project health and
defect prediction (Table~\ref{tab:datasets}), and the algorithms we selected cover five 
families and six of the seven assumptions an SE practitioner would plausibly try
(Table~\ref{tab:inventory}), given a reading of the current SE
optimization literature.
\item
When we exclude algorithms, we take care to justify why they are
excluded   (see section~\ref{ass});
\item
We publish all our scripts and data on-line so other researchers can apply our methods to their data. 
\end{itemize}
The opposite of using too little test data is using too much. 
A related threat is that the guide could be overfit to these tasks. \emph{Mitigation:}
the guide proposed in this paper is validated under stratified cross-validation against a hindsight oracle
on held-out tasks (Table~\ref{tab:guideval}), not on the data it was fit to.

\textbf{Conclusion validity.} Differences between methods could be noise. To mitigate this, every cell is repeated over 20 seeds, all rankings use Scott-Knott
clustering gated by a bootstrap test and a Cliff's delta effect-size threshold (Section~\ref{sec:rig}), so reported
wins survive both significance and effect-size screens.

\section{Conclusion}
\label{sec:conclusion}
Search-based software engineering offers practitioners an overwhelming menu of
black-box optimizers and almost no budgeted rule for choosing among them. Recasting
that menu as a set of bets on algorithm assumptions, we organized 20 optimizers into an
assumption-indexed tournament and ran it over 106 SE tasks at four budgets, roughly
14{,}000 CPU hours. The tournament yields no single optimizer. Instead it exposes a
\emph{boundary}: the winner migrates from a fast-warming geometric active learner at
tight budgets to model-free differential evolution once the budget grows, and the single
best optimizer differs from its own $B{=}30$ choice on up to \textbf{50\%} of tasks by
$B{=}200$, with \textbf{58\%} switching at least once across the four budgets we tested.

The standard way to sidestep such a study, which is predicting the winner from instance
clustering metrics, fails at SE budgets (\textbf{44.2\%} accuracy). 

Hence, we propose another guide based on some cheap-to-obtain attributes (objective conflict,  input-space shape, labeling
budget). This second guide scores as well as a hindsight oracle on
\textbf{74.2\%} of held-out tasks at near zero probe cost. Our takeaway for practitioners:
\begin{finding}
 Treat optimizer selection in SE as a budget-dependent matrix: anchor the choice on cheap structural signals, and leave expensive instance clustering methods   behind.
\end{finding}

The research message is larger. The boundary reframes optimizer selection as a
\emph{scheduling} problem rather than a one-shot pick. The migration we measured (a
geometric learner early, model-free evolution late) suggests a \emph{warm handoff}:
spend the first evaluations with EZR to locate good regions cheaply, then hand its best
points to DE as the budget accrues. This is adjacent to multi-fidelity methods such as
Hyperband, BOHB, and DEHB~\cite{li2018hyperband,falkner2018bohb,awad2021dehb}, but the
axis differs: those schedule \emph{fidelity} within one algorithm, whereas our boundary
schedules the \emph{algorithm}, and with it the landscape assumption. 

For future work, we suggest   four  directions:
\begin{enumerate}
\item
Build the meta-scheduler the boundary
implies: a controller that switches assumptions at the measured crossover rather than
fidelities within a fixed assumption. 
\item Test whether that boundary survives richer
methods, such as causal, multi-fidelity, high-dimensional Bayesian, and LLM-driven
optimizers~\cite{chen2026promisetune,falkner2018bohb,eriksson2019scalable,zhang2025using}, each encoding an assumption our seven omit, and each may move the budget crossover or leave it in
place. 
\item
Sharpen the boundary itself: with only four budget points the EZR$\to$DE
crossover is real but fuzzy, so
a continuous-budget characterization, or a blend rule near the crossover could be the next
refinement.
\item
Extend the vocabulary of cheap-to-obtain attributes. 
\end{enumerate}
Beyond these, our guide has a natural role in LLM-driven and agentic search. As a
\emph{prior}, it offers a cheap, auditable first choice. As a \emph{baseline}, it sets a
bar any added sophistication must clear: matching a hindsight oracle on three of four
tasks from two cheap-to-obtain attributes is the number to beat.

\balance
\bibliographystyle{IEEEtran}
\bibliography{refs}

\end{document}